\documentclass[final,3p,times,twocolumn]{elsarticle}
 \biboptions{comma,sort&compress}
\usepackage{here}
 \usepackage{graphicx}
  \usepackage{epsfig}

\def\nin{\noindent}
\def\beq{\begin{equation}}
\def\eeq{\end{equation}}
\def\bea{\begin{eqnarray}}
\def\eea{\end{eqnarray}}
\def\enq{\end{equation}}
\def\beqa{\begin{eqnarray}}
\def\enqa{\end{eqnarray}}

\def\GeV{\nobreak\,\mbox{GeV}}

\def\lb{\label}
\def\rag{\rangle}
\def\lag{\langle}

\journal{Nuc. Phys. (Proc. Suppl.)}

\begin{document}

\begin{frontmatter}



\title{Investigating different structures for the $X(3872)$}

 \author[label1]{Stephan Narison}
\ead{snarison@yahoo.fr}
 \author[label2]{Fernando S. Navarra}
\ead{navarra@if.usp.br}
 \author[label2]{Marina Nielsen\corref{cor1}}
\ead{mnielsen@if.usp.br}
\cortext[cor1]{Speaker}
  
\address[label1]{Laboratoire
de Physique Th\'eorique et Astroparticules, CNRS-IN2P3 \& Universit\'e
de Montpellier II, \\
Case 070, Place Eug\`ene
Bataillon, 34095 - Montpellier Cedex 05, France.}

  \address[label2]{Instituto de F\'{\i}sica, Universidade de S\~{a}o Paulo, 
C.P. 66318, 05389-970 S\~{a}o Paulo, SP, Brazil}


\begin{abstract}
\noindent
Using the QCD spectral sum rule approach we investigate different currents
with $J^{PC}=1^{++}$, which could be associated with the
$X(3872)$ meson. Our results indicate that, with a four-quark or 
molecular structure, it is
very difficult to explain the narrow width of the state unless the
quarks have a special color configuration. 
\end{abstract}




\end{frontmatter}


\section{Introduction}
\nin
In the last years, several new observations on hadron states came from a 
variety of facilities. These include, in particular, low-lying excitations 
of the charmonium states, like the $X(3872)$ \cite{BELLE}, the $Y(4260)$ 
\cite{babary}, the $Y(3940)$ \cite{belley} the $Z(3930)$ \cite{bellez} 
the $X((3940)$ \cite{bellex} and many others with higher masses. 
Among these new hadron states, some of them are good candidates for exotic
structures like: hybrid mesons, four-quark states, cusp, glueball or 
meson-meson molecules \cite{Swanson}. 
In this work we use QCD spectral sum rules (QSSR/QCDSR) (the Borel/Laplace Sum
Rules \cite{svz,rry,SNB}) to study the two-point functions of the axial
vector meson, $X(3872)$, assumed to be a four-quark state. Experimental
information about the $X(3872)$ and previous uses of the QCDSR to study the 
$X(3872)$ meson can be found in \cite{rev}.

The idea of mesons as four-quark states is not new. Indeed, even Gell-Mann
in his first work about quarks had mentioned that mesons could
be made out of $(q\bar{q}),~(qq\bar{q}\bar{q})$, ... \cite{gell}. The 
well known example of applying the idea of four-quark states for mesons is 
for the light scalar mesons (the isoscalars $\sigma(500),~f_0(980)$, the 
isodublet $\kappa(800)$  and the isovector $a_0(980)$) \cite{jaffe,cloto}. 
In the four-quark scenario,  mesons can be considered as 
states of diquark-antidiquark pairs, or molecular (two-meson) bound states.
The idea of diquarks is almost as old as 
QCD \cite{diq} and have been the subject of intense study by many theorists
\cite{exo}. As pointed out in \cite{wil}, diquarks can form bound states, 
which can be used as degrees of freedom in parallel with quarks themselves.
For a diquark-antidiquark state, the color singlet can be 
obtained with
the diquark-antidiquark pairs in a $\bar{3}-3$ or $6-\bar{6}$ color 
configuration.
In  previous calculations  some of us have considered the $X(3872)$ as being 
a tetraquark state in the $\bar{3}-3$ color configuration \cite{mnnr} and
in a molecular $D^*D$ configuration \cite{x24}. The currents used in these
studies were
\beqa
j_\mu^{(3)}&=&{i\epsilon_{abc}\epsilon_{dec}\over\sqrt{2}}[(q_a^TC\gamma_5c_b)
(\bar{q}_d\gamma_\mu C\bar{c}_e^T)
\nonumber\\
&+&(q_a^TC\gamma_\mu c_b)
(\bar{q}_d\gamma_5C\bar{c}_e^T)]\;,
\label{field3}
\enqa
for the tetraquark state in the $\bar{3}-3$ color configuration and
\beqa
j^{mol}_{\mu}(x) & = & {1 \over \sqrt{2}}\Big{[}
\left(\bar{q}_a(x) \gamma_{5} c_a(x)
\bar{c}_b(x) \gamma_{\mu}  q_b(x)\right)\nonumber\\
&&- \left(\bar{q}_a(x) \gamma_{\mu} 
c_a(x)\bar{c}_b(x) \gamma_{5}  q_b(x)\right)
\Big{]},
\lb{fieldm}
\enqa
for the molecular $D^*D$ configuration. In the above equations $a,~b,~c,~...$
 are color indices, $C$ is the charge conjugation matrix and $q$ denotes a 
$u$ or $d$ quark.

The results obtained with these two currents are very similar
and  the deviations between the two QCDSR
results in the allowed Borel window are smaller than 0.01\% \cite{rev}.

Another possible current with $J^{PC}=1^{++}$ and the 
diquark-antidiquark in the color sextet configuration is
\beqa
j_\mu^{(6)}&=&{i\over\sqrt{2}}[(q_a^TC\gamma_5\lambda^S_{ab}c_b)
(\bar{q}_d\gamma_\mu C\lambda^S_{de}\bar{c}_e^T)~+
\nonumber\\
&+&(q_a^TC\gamma_\mu
\lambda^S_{ab} c_b)(\bar{q}_d\gamma_5C\lambda^S_{de}\bar{c}_e^T)]\;,
\label{field}
\enqa
where  $\lambda^S$ stands for the 
six symmetric Gell-Mann matrices
$\lambda^S=\left(\lambda_0,~\lambda_1,~\lambda_3,~\lambda_4,~\lambda_6,
~\lambda_8\right)$.

In the QCDSR approach, the mass of the particle can be determined
by considering the two-point correlation function
\beq
\Pi_{\mu\nu}(q)=i\int d^4x ~e^{iq.x}\lag 0 |T[j_{\mu}(x)
j^\dagger_{\nu}(0)]|0\rag.
\lb{2po}
\enq

In the phenomenological side, Eq.~(\ref{2po}) can be written as
\beq
\Pi^{phen}_{\mu\nu}(q^2)={\lambda^2\over m_{X}^2-q^2}\left(-g_{\mu\nu}
+{q_\mu q_\nu\over m_X^2}\right)+\cdots\;,
\lb{phex}
\enq
for the $X$ meson represented by the currents in Eqs.~(\ref{field3}),
(\ref{fieldm}) and (\ref{field}). In Eq.~(\ref{phex}),
the dots denote higher resonance contributions that will be 
parametrized, as usual, through the introduction of the continuum threshold
parameter $t_c$, and where we have introduced the meson-current
coupling, $\lambda$, through the parametrization
$\lag 0 | j_\mu^q|X(q)\rag =\lambda\epsilon_\mu(q)$.

In the OPE side, we work at leading order and consider condensates up to 
dimension six. We can write the correlation function in the OPE side in 
terms of a dispersion relation. Generically, we have
\beq
\Pi^{OPE}(q^2)=\int_{4m_c^2}^\infty ds {\rho(s)\over s-q^2}\;,
\lb{ope}
\enq
where the spectral density is given by the imaginary part of the correlation
function: $\rho(s)={1\over\pi}\mbox{Im}[\Pi^{OPE}(s)]$.

After making an inverse-Laplace (or Borel)
transform on both sides, and transferring the continuum contribution to
the OPE side, the sum rule can be written as
\beq
\lambda^2e^{-(m_{X}^{(N)})^2\tau}=\int_{4m_c^2}^{t_c}ds~ e^{-s\tau}~
\rho^{(N)}(s)\;,
\lb{sr}
\enq
where $N=3,mol,6$ are related with the currents in Eqs.~(\ref{field3}), 
(\ref{fieldm}) and (\ref{field}) respectively, and the complete expressions 
for $\rho^{(3)}(s)$, $\rho^{(mol)}(s)$ and $\rho^{(6)}(s)$ are given in 
refs.~\cite{mnnr,x24,nnn}.

Using the QCDSR method, one usually estimates the hadron mass from the ratio
\beq
{\cal R}_N={\int_{4 m_c^2}^{t_c}ds ~e^{-s\tau}~s~\rho^{(N)}(s)\over\int_{4
m_c^2}^{t_c} ds ~e^{-s\tau}~\rho^{(N)}(s)}\sim (m_X^{(N)})^2\;.
\lb{m2}
\enq
However,  in this work, instead of evaluating the meson mass through 
Eq.~(\ref{m2}), we will consider the double ratios (DR) of the sum rules 
(DRSR) \cite{nnn}
\beq
r_{6/3}=\sqrt{{\cal R}_{6}\over{\cal R}_{3}}\simeq {m_X^{(6)}\over m_X^{(3)}},
\lb{dr3}
\enq
and
\beq
r_{mol/3}=\sqrt{{\cal R}_{mol}\over{\cal R}_{3}}\simeq {m_X^{(mol)}
\over m_X^{(3)}},
\lb{drm}
\enq

This quantity has the advantage to be less sensitive to the perturbative 
radiative corrections and continuum contributions than the single ratio in 
Eq.~(\ref{m2}). Therefore, we 
expect that our results obtained  to leading order 
in $\alpha_s$ will be quite accurate.

\section{QCDSR Results}
\nin

\begin{figure}[hbt] 
\centerline{\includegraphics[width=7.cm]{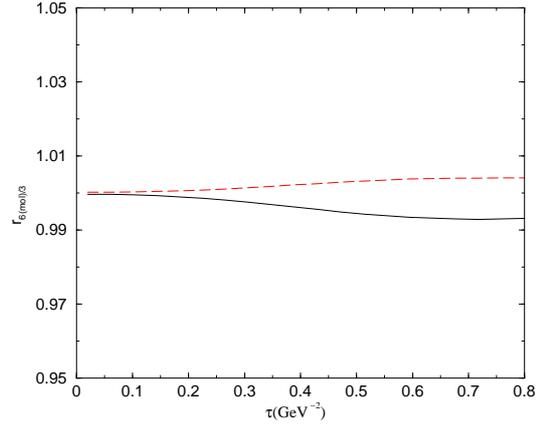}}
\caption{\small The double ratio $r_{6/3}$ (solid line) and 
$r_{mol/3}$ (dashed line) as a function of $\tau$ for $\sqrt{t_c}
=4.15~\GeV$.}
\label{fig1} 
\end{figure} 

Fixing $\sqrt{t_c}=4.15~\GeV$, the value obtained in ref.~\cite{mnnr}, we show
in Fig.~\ref{fig1} the $\tau$-behaviour of  $r_{6/3}$ (continuous line) and
$r_{mol/3}$ (dashed line), using the QCD parameters presented in Table 
\ref{tab:param} \cite{nnn}.

 \vspace*{-0.5cm}
{\scriptsize
\begin{table}[hbt]
\setlength{\tabcolsep}{0.5pc}
 \caption{\small   QCD input parameters.}
    {
\begin{tabular}{cc}
&\\
\hline
Parameters&Values \\
\hline
$\Lambda(n_f=4)$& $(324\pm 15)$ MeV \\
$\hat \mu_d$&$(263\pm 7)$ MeV\\
$m_0^2$&$(0.8 \pm 0.1)$ GeV$^2$\\
$\lag\alpha_s G^2\rag$& $(6\pm 1)\times 10^{-2}$ GeV$^4$\\
$\rho\alpha_s\lag \bar dd\rag^2$& $(4.5\pm 0.3)\times 10^{-4}$ GeV$^6$\\
$m_c$&$(1.26\sim1.47)$ GeV \\
\hline
\end{tabular}
}
\label{tab:param}
\end{table}
}

From Fig.~\ref{fig1} one can notice that 
the results are very stable against the $\tau$-variation in a large range for 
$\tau\leq 0.8$ GeV$^{-2}$ . We deduce
\beq
 r_{6(mol)/3}= 1.00,
 \lb{eq:63}
\enq
with a negligible error, which shows that, from a QCD spectral sum rules 
approach, the $X(3872)$ can be equally described by the currents in 
Eqs.~(\ref{field3}), (\ref{fieldm}) and (\ref{field}).

The problem with these three currents is that a QCDSR study of the decay width
of the decays $X\to J/\psi +3\pi$ and $X\to J/\psi+ 2\pi$ \cite{nnn,NN},
gives 
\beq
\Gamma(X\to J/\psi +n\pi)\vert_{3,6,mol}\approx 50~{\rm MeV}~,
\lb{eq:gam3m}
\enq
which is too big compared with the data upper limit \cite{BELLE}
\beq
 \Gamma(X\to {\rm all}) \leq  2.3~{\rm MeV}~.
 \label{eq:bound}
 \enq

The QCDSR  study of the $X-\psi-V$ couplings, where the $2\pi$ and $3\pi$ 
can be assumed to come from the vector mesons $\rho$ and $\omega$, is based
on the three-point function
\beqa
\Pi^{\mu\nu\alpha}(p,p',q)&=& \int d^4x~d^4y ~e^{i(p'x+qy)}\times
\nonumber\\
&&\langle 0\vert{\cal T}J_\psi^\mu(x)J_V^\nu(y)J_X^{\alpha\dagger}(0)\vert 0
\rangle
\enqa
associated to the $J/\psi$-meson $J_\psi^\mu$, vector mesons 
$J_V^\nu$ and to the $X$-meson $J_X$.

\begin{figure}[hbt] 
\centerline{\epsfig{figure=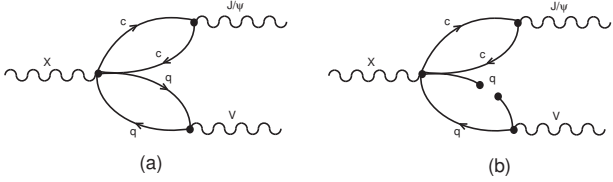,width=8.cm}}
\caption{\small{Lowest order and lowest dimension vertex diagrams 
contributing to the $X$-width for the diquark currents in 
Eqs. (\ref{field3}), (\ref{fieldm}) and (\ref{field}).}}  
\label{vertex1} 
\end{figure}

In the case of the three $X$-currents ($\bar 3-3$, $6-6$ tetraquarks and 
molecule) discussed above, the lowest order and lowest dimension 
contributing QCD diagrams are shown in Fig. \ref{vertex1}. 

\begin{figure}[hbt] 
\centerline{\includegraphics[height=40mm]{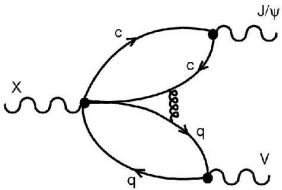}}
\caption{\small Lowest order vertex diagrams due to one gluon exchange.}
\label{onegluon} 
\end{figure}

As a matter of fact, a large partial decay width for the decay $X\to J/\psi
V$ should be expected in this case.
The initial state already contains all the four quarks needed for the decay,
and there is no selection rules prohibiting the decay. Therefore, the decay
is super-allowed through the fall-apart mechanism represented by the
diagrams in Fig. \ref{vertex1}. To avoid such fall-apart mechanism,
the leading order contribution to the three-point function should be
due to one gluon exchange, as shown in Fig. \ref{onegluon}. 

A possible current, for which the leading order contribution to the 
three-point function is due to one gluon exchange, as in Fig. \ref{onegluon},
 is  the $\lambda$-molecule $J/\psi-\pi$-like current
\beq
j_\mu^{{mol}\vert^{\lambda}}  =  
\left(\bar{c}\lambda^a\gamma_{\mu}c\right)
\left(\bar{q}\lambda_a  \gamma_5 q\right)
\lb{eq:curr5}
\enq
where $\lambda_a$ is the colour matrix. 

\begin{figure}[h]
\centerline{\epsfig{figure=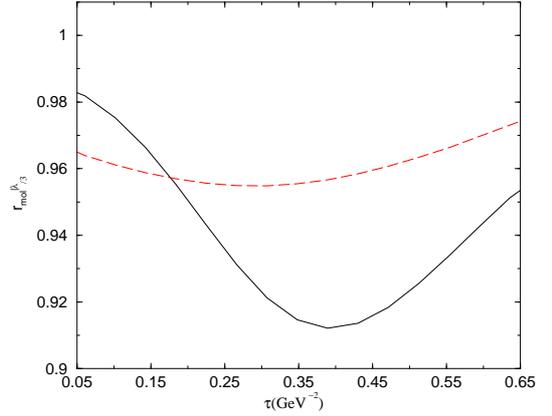,width=7.0cm}}
\caption{\small{The double ratios $r_{\lambda/3}$  as a function of $\tau$ 
for $\sqrt{t_c}= 3.9$ GeV (red dashed line) and 5 GeV (black continuous line)
. }}
\label{fig4}
\end{figure}

The corresponding spectral functions for this current are also given in
Ref.~\cite{nnn}. In Fig.~\ref{fig4}, we show the ratio of the 
$\lambda$-molecule over the tetraquark $3-\bar{3}$ one.

In this case we get
\beq
r_{{mol|^\lambda}/3}=0.96\pm 0.03,
\label{eq:ratiolam}
\enq
where the errors come from the stability regions and $m_c= 1.26$ to 1.47 GeV 
(running MS and pole mass). This  result suggests that  
the $X(3872)$ can have a large {\it $\lambda$-molecule} component rather than 
a $6-6$, $\bar 3-3$ tetraquark or an (usual) $D-D^{(*)}$ molecule one.  If 
this is the case, one expects, from Eq. (\ref{eq:ratiolam}), that the 
tetraquarks and $D-D^{(*)}$ molecule states are in the range of 
3910-4160 MeV and with broader widths. 

\section{Conclusions}
\nin

In conclusion,
we have studied the mass of the $X(3872)$ using double ratios of sum rules. 
We found that the different proposed substructures ($\bar 3-3$ and $6-6$ 
tetraquarks and molecules) lead to the same mass predictions within the 
accuracy of the method [see Eq.  (\ref{eq:63})], 
indicating that the predictions of the $X$ meson mass is not sufficient for 
revealing its nature.

Among these different proposals, the only eventual possibility which can lead 
to a narrow width X(3872) is the  choice of $\lambda$-molecule-$J/\psi$-like 
current given in Eq. (\ref{eq:curr5}). If this is the case, then the 
tetraquarks and $D-D^{(*)}$ molecule states are in the range of 3910-4160 
MeV and with broader widths. Some further tests of this proposal are welcome.

\section*{Acknowledgements}
\nin
M. Nielsen would like to thank the LPTA-Montpellier for the hospitality
where this work has been initiated. This work has 
been partly supported by the CNRS-FAPESP within the QCD program,  by  
CNPq-Brazil and by the IN2P3-CNRS within QCD program in Hadron Physics. 





\begin{thebibliography}{999}
\vspace*{-0.25cm}
\bibitem{BELLE} Belle Coll.,  S.-L. Choi  {\it et al.},    
Phys. Rev. Lett. {\bf 91}, 262001 (2003).
\bibitem{babary} BABAR Coll., B. Aubert {\it et al.},  Phys. Rev. Lett. 
{\bf 95}, 142001 (2005).
\bibitem{belley} Belle Coll.,
  S.-K. Choi {\it et al.}, hep-ex/0505037, Phys. Rev. Lett. {\bf 94}, 182002 
(2005).
\bibitem{bellez} Belle Coll.,
 S. Uehara {\it et al.}, Phys. Rev. Lett. {\bf 96}, 082003 (2006).
\bibitem{bellex} Belle Coll.,
 K. Abe {\it et al.}, hep-ex/0507019.
\bibitem{Swanson} for a review see E.~S.~Swanson,
  Phys.\ Rept.\  {\bf 429}, 243 (2006).
\bibitem{svz} M.A. Shifman, A.I. and Vainshtein and V.I. Zakharov,
Nucl. Phys. {\bf B147}, 385 (1979).

\bibitem{rry} L.J. Reinders, H. Rubinstein and S. Yazaki, Phys. Rept. 
{\bf 127}, 1 (1985). 

\bibitem{SNB} For a review and references to original works, see
e.g., S. 
Narison, {\it QCD as a theory of hadrons,
Camb. Monogr. Part. Phys. Nucl. Phys. Cosmol.} {\bf 17}, 1-778
(2002) 
[hep-h/0205006]; {\it QCD
spectral sum rules ,  World Sci. Lect. Notes Phys.} {\bf 26}, 1-527
(1989);
{ Acta Phys. Pol.} {\bf B26} (1995) 687; { Riv. Nuov. Cim.} {\bf 10} 
(1987) 1; { Phys. Rept.} {\bf 84}, 263 (1982).

\bibitem{rev} M. Nielsen, F.S. Navarra, S.H. Lee, arXiv:0911.1958.

\bibitem{gell} M. Gell-Mann, Phys. Lett. {\bf8}, 214 (1964).
\bibitem{jaffe} R.L. Jaffe, { Phys. Rev.} {\bf D15}, 267, 281 (1977);
{\bf D17}, 1444 (1978).
\bibitem{cloto} For a review see F.E. Close and N.A. T\"ornqvist, 
{ J. Phys.} {\bf G28}, R249 (2002).
\bibitem{diq} M. Ida and R. Kobayashi, Prog. Theor. Phys. {\bf36}, 846 (1966).
\bibitem{exo} For a review see, e.g., R.L. Jaffe,
  Phys.\ Rept.\  {\bf 409}, 1 (2005).
\bibitem{wil} F. Wilczek, hep-ph/0409168.

\bibitem{mnnr} R.D. Matheus, S. Narison, M. Nielsen, J.M. Richard,
Phys. Rev. {\bf D75}, 014005 (2007).

\bibitem{x24} R. D. Matheus, F. S. Navarra, M. Nielsen, C. M. Zanetti,
Phys. Rev. {\bf D80}, 056002 (2009).

\bibitem{nnn} S. Narison, F.S. Navarra,  M. Nielsen, arXiv:1006.4802,

\bibitem{NN} F. Navarra and M. Nielsen, Phys. Lett. {\bf B639},272 (2006).


\end{thebibliography}








\end{document}